# A Model of Proactive Safety Based on Knowledge Graph

He Wen

Department of Civil and Environmental Engineering,

University of Alberta, Edmonton, AB T6G 2E1, Canada

Correspondence author: hwen7@ualberta.ca

**Abstract**

In contemporary safety management, despite the abundance of safety data gathered from routine operation tasks and safety management activities, actions cannot prevent all accidents effectively due to a lack of effective utilization of these data as safety knowledge. To bridge this gap, this paper proposes a hybrid proactive safety model integrating data-driven and knowledge-driven approaches. The model comprises three main steps: proactive safety actions to generate safety data, data-driven approaches to mine safety data, and knowledge-driven approaches to depicting risk knowledge graphs. Application of this model to a continuous stirred tank reactor (CSTR) scenario demonstrates its efficacy in identifying and addressing safety issues proactively. The results demonstrate the effectiveness and practicality of the proposed proactive safety model, suggesting its endorsement within both academic and industrial applications.

# 1 Introduction

In the digital era, data stands as a crucial component and resource. Safety data, in particular, manifests in diverse forms, ranging from numerical data by measurement and simulation to textual data by surveying and expert involvement (Wen et al., 2022). They emerge from routine operation tasks and safety management activities like safety inspections, tests, audits, and structural hazard identification processes, even exhibiting business intelligence (Nakhal A et al., 2021). Despite the regularity of safety activities, accidents and incidents persist. It is noted that these efforts gather substantial safety data, but have not been discovered as structural safety knowledge.

Not only do raw data face such problems, but even the safety information obtained through structured hazard identification is difficult to directly guide practice. In the realm of hazard identification, numerous techniques have been proposed and employed, for example, Failure Modes and Effects Analysis (FMEA), Hazard and Operability Study (HAZOP), and Job Hazard Analysis (JHA). They usually require high-intensity intellectual labor. Though the outcomes of these efforts appear promising, addressing the uncertainties and probabilities of consequences (Pasman and Rogers, 2018), it seems that they often manifest as mere static tables and sheets (Table 1), swiftly fading into obscurity.

Table 1: Methods of hazard identification and corresponding outputs.

| Method | Output |
| --- | --- |
| Checklist | A table of hazards and conditions |
| Brainstorming | A table of hazards, causes, consequences, and actions |
| FMEA | A table of hazards, causes, consequences, and actions |
| HAZOP | A table of hazards, causes, consequences, and actions |
| JHA | A table of activities, hazards, causes, consequences, and actions |
| Fault Tree Analysis (FTA) | A sheet of hazards; probability of the accident |
| Event Tree Analysis (ETA) | A sheet of consequences and their probabilities |
| Bowtie Analysis (BT) | A sheet of hazards, consequences, and probabilities |

For an accident, frontline operators, rather than safety engineers equipped with this safety knowledge, typically act as the initial responders. This underscores a potential failure in leveraging existing safety knowledge to predict, diagnose, and implement appropriate measures. Therefore, the question arises: How can we harness the data and information generated by these routine tasks to unearth safety knowledge and inform practical applications?

This has always been lacking. One piece of evidence is that according to US OSHA's top 10 most frequently cited standards for fiscal year 2022, the second most violation is about hazard communication, and this is also shown in similar statistics from different perspectives (Wang et al., 2021). Although hazard here generally refers to hazardous chemicals, which is one type of hazard, it can be seen that most workers have not received enough hazard information. Some studies have pointed out that even if it only focuses on knowledge communication of hazardous chemicals, there are still challenges (Koshy et al., 2015).

Moreover, comparing safety practice with academia, there may have been differences in focus. Scholars usually stress how to improve hazard identification technology, reflecting on the shortcomings of existing technology and future improvement (Lee et al., 2019; Pasman et al., 2018). However, there is insufficient research to elucidate the process of transforming hazard identification results into safety knowledge, not to mention research on how to guide the incorporation of safety knowledge into standard operating procedures (SOP). Nevertheless, deriving safety insights from accidents has perennially been a focal point of academic endeavors, whether through traditional experiential analysis, accident text mining, or comparing and integrating with hazard identification methods (Paltrinieri et al., 2012). These studies have not yet risen to the fields of knowledge discovery and wisdom discovery.

On the other hand, academics also recognize the significance of safety knowledge. For example, one survey demonstrates that safety knowledge significantly influences both safety compliance and safety participation. Typically, inefficient systems for storing and transferring knowledge hinder the workforce's ability to access vital information necessary for addressing urgent safety issues (Hallowell, 2012). Studies also examine the organizational climate for safety training transfer as a moderator of the relationship between safety knowledge and safety performance (Smith-Crowe et al., 2003). Moreover, there have been some explorations in applying knowledge graphs to the safety field. For instance, one study integrates a safety library in building information modeling (BIM) (Hossain et al., 2018). There are also some preliminary studies on knowledge graphs based on natural language processing (NLP) and ontology (Chua and Goh, 2004; Lu et al., 2015a, 2015b; Zhang et al., 2015). These attempts help to build a safety knowledge repository but have not yet proposed how to apply it to safety practices. Moreover, current large language models (LLMs) based on generative AI can produce some forms of knowledge, while safety knowledge, which is highly structured, cannot be directly and reliably accessed using LLMs. Therefore, this requires further building a proactive safety model based on safety knowledge.

Hence, leveraging knowledge graphs and data mining might provide a potential solution. Presently, within the realm of artificial intelligence (AI), data-driven technology (e.g., deep learning) and knowledge-based technology (e.g., knowledge graph) represent the two predominant directions (Fig. 1). In prior research, data mining techniques have been extensively utilized in fault diagnosis applications (Alauddin et al., 2018). However, it is crucial to note that this primarily remains at the stage of safety information acquisition. This is due to the scarcity of available information within the normal data to infer anomalies, with abnormal data being typically sparse (Liu and Feng, 2022). As a result, the focus tends to be on restoring normalcy rather than identifying the root cause of anomalies for safety purposes. In addition, Knowledge graphs resemble Bayesian networks in structure but differ as they do not propose and manipulate probability values (Wen et al., 2022). Instead, they concentrate on storing and displaying structured knowledge, effectively serving as a practical bridge between academic theory and real-world application.

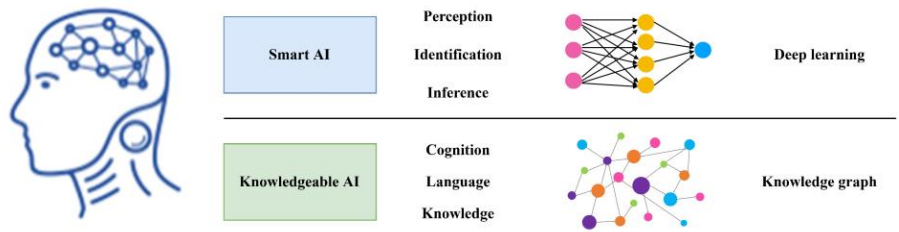

Fig. 1. Two types of AI.

Therefore, this study will propose a hybrid proactive safety model based on data-driven and knowledge-driven approaches, aiming to address the following questions:

i. How can we harness the data and information generated by routine tasks to unearth safety knowledge?
ii. How can we help workers acquire sufficient safety knowledge to deal with abnormalities and emergencies?

To provide readers with a clear roadmap of the paper, it is important to note the following: Section 2 outlines the proposed proactive safety model; Section 3 illustrates its application to the continuous stirred tank reactor (CSTR); Section 4 delves into the discussion, while Section 5 presents notable conclusions.

## 2 The proactive safety model

### 2.1 General description

The proposed model of proactive safety includes 3 major steps (Fig. 2).

**Step 1**: Workers conduct proactive safety actions (e.g., inspection) and generate data indicating safety issues.

**Step 2**: Utilize data mining techniques to explore the data and discover the information and knowledge.

**Step 3**: Utilize knowledge graph to depict risk knowledge graph, then export risk warning and countermeasure.

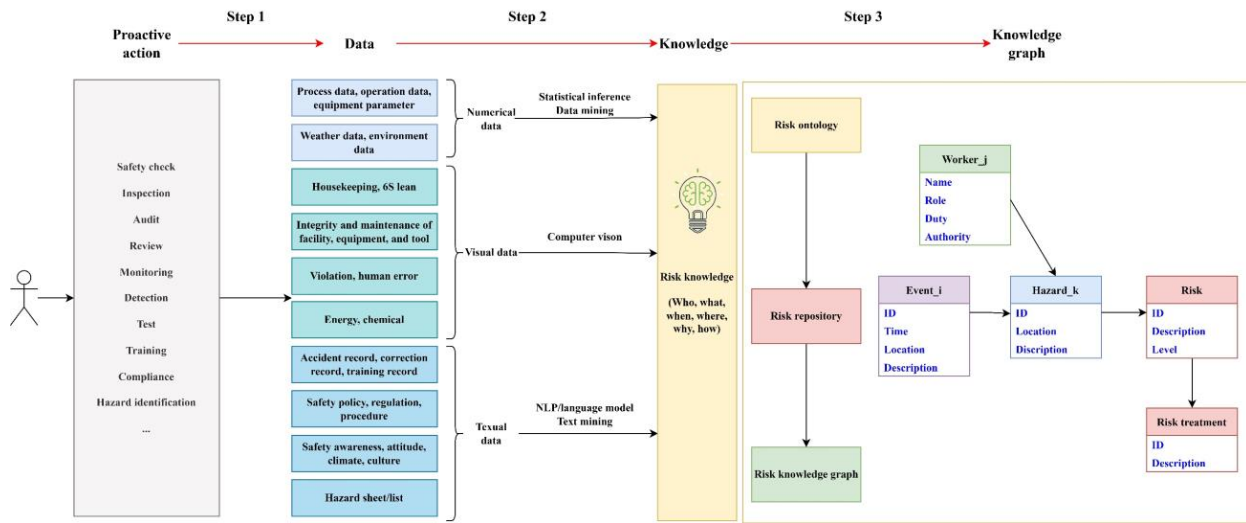

Fig. 2. Model of proactive safety.

### 2.2 Step 1: Proactive safety actions to generate safety data

Proactive safety requires actions before the accidents, generally, they include safety checks before work, regular inspection, ISO management system audit, management review, dynamic monitoring, detection, test, training, education, regulatory compliance, and structural systematic hazard identification and risk assessment.

Subsequently, these actions will generate numerical data, textual data, and visual data. These data may indicate abnormalities, like fault, failure, deviation, gap, conflict, near miss, mishap, and incident. Direct inference can help to obtain safety information, while further knowledge needs to be explored with data techniques. The procedure is shown in Fig. 3.

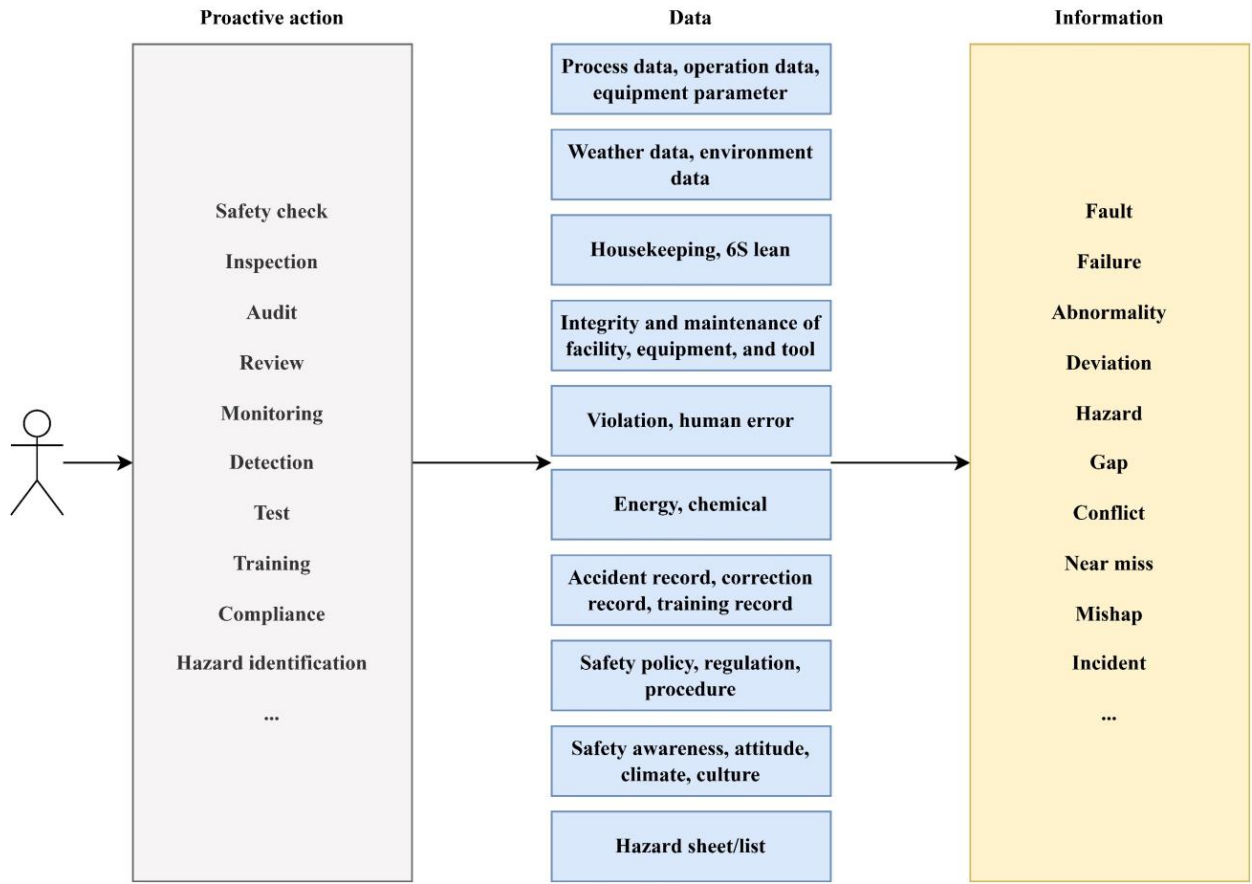

Fig. 3. From proactive action to data.

## 2.3 Step 2: Data-driven approaches to mine safety data

Developing the safety data/information to knowledge requires hybrid AI technologies. For numerical data, statistical approaches and machine learning algorithms are widely used for fault diagnosis and detection, for example, principle component analysis (PCA) and other multivariate statistics-based process monitoring (MSPM). For visual data, image and video recognition based on computer vision has also been employed for safety purposes. For textual data, natural language processing and large language models (e.g., Generative Pre-trained Transformer (GPT)) are also in experiments for safety uses. All these techniques can mine the data and produce safety knowledge which can be generally expressed with 5W1Y (who, what, when, where, why, how), describing the mechanism of risks (Fig. 4).

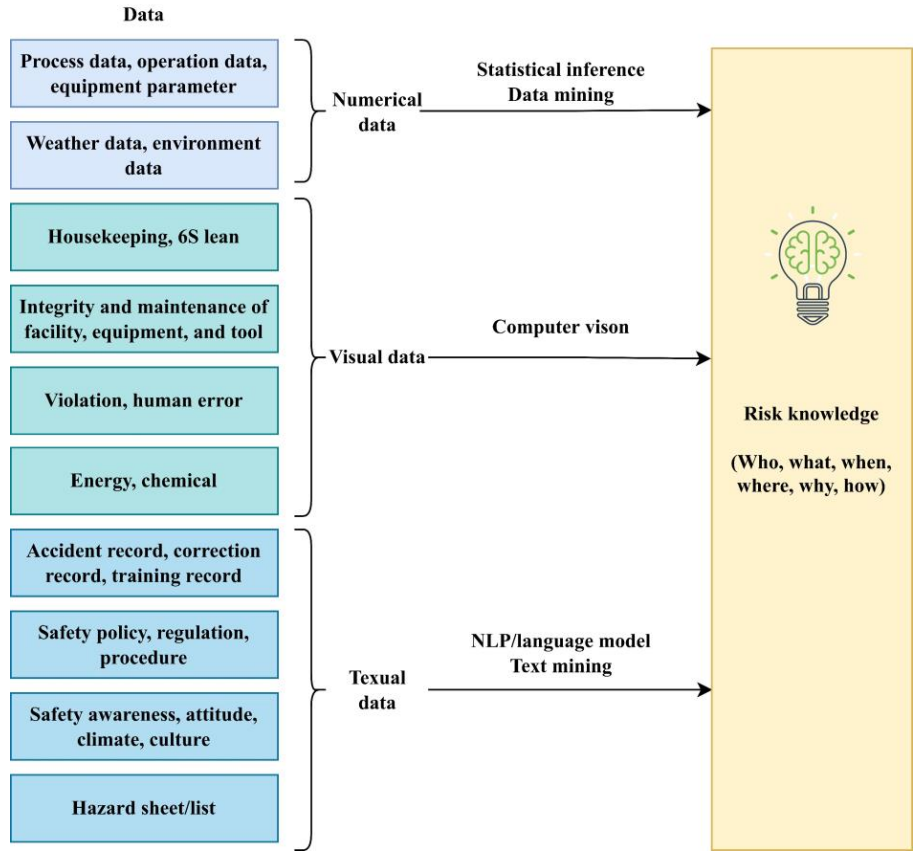

Fig. 4. From safety data to knowledge.

## 2.4 Step 3: Knowledge-driven approaches to depict risk graph

Long descriptive expressions of safety knowledge cannot be applied in the field for the workers directly. Therefore, it is necessary to establish a risk knowledge graph as an AI agent, assisting the worker all the time. First, the approach is to use the concept of ontology to describe the relations among the key elements of the risk knowledge (worker, event, hazard, risk, risk treatment), which is to describe and instead of 5W1Y by data entity and structure. Thus, the AI can also understand the risk knowledge in natural language (Fig. 5).

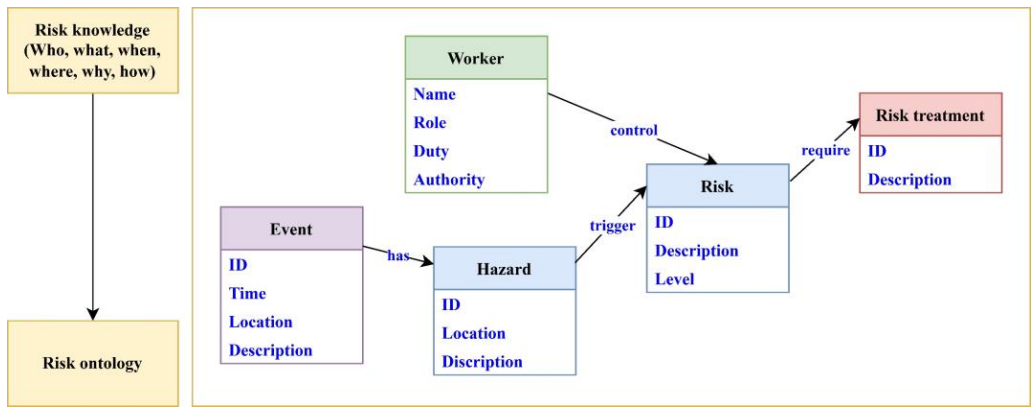

Fig. 5. Risk ontology.

Second, the task is to consider how to store the knowledge and sort them well. Here define three cores: event, worker, and hazard; then explore the relations between events and connect all events together in the database (Fig. 6). Generally, there are four types of relationships in events: causal, sequential, hierarchical, and simultaneous (Knez and Žitnik, 2023; Liu et al., 2021). Similarly, the process is repeated to analyze "worker" (Fig. 7) and "hazard" (Fig. 8).

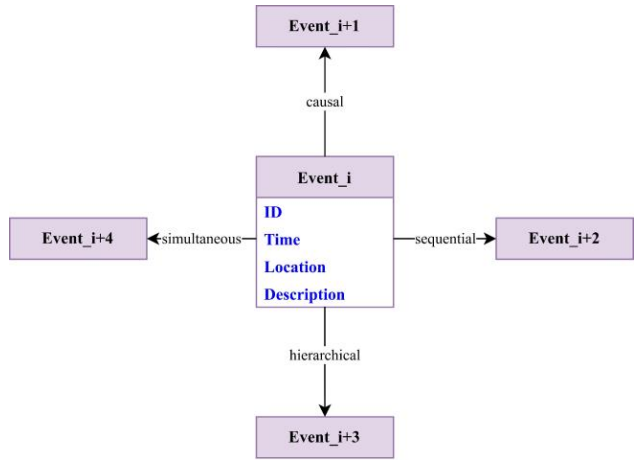

Fig. 6. Core of event in risk repository.

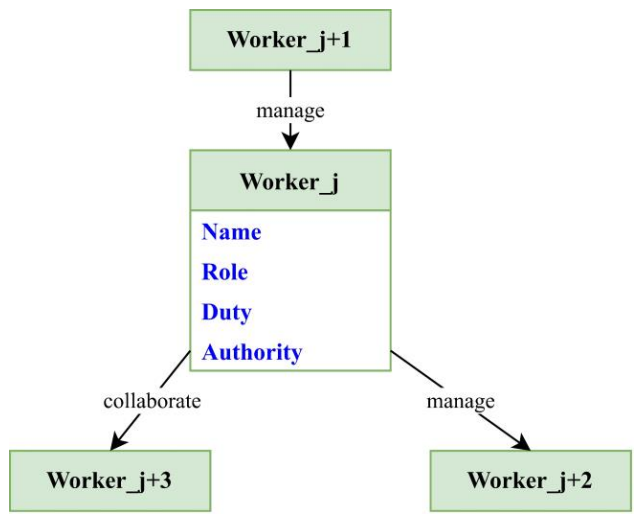

Fig. 7. Core of worker in risk repository.

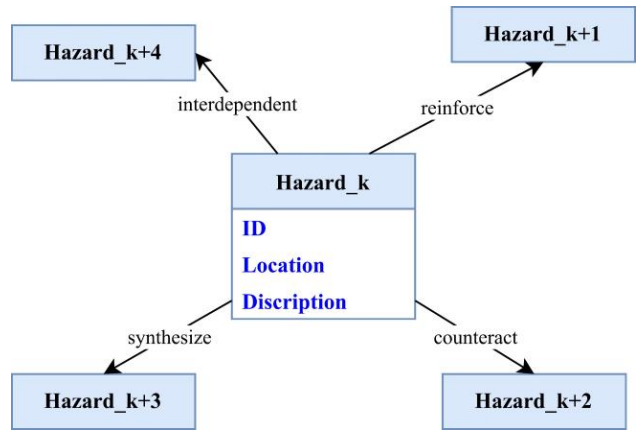

Fig. 8. Core of hazard in risk repository.

Third, unite the risk ontology and risk repository to organize a big picture of the risk knowledge graph (Fig. 9).

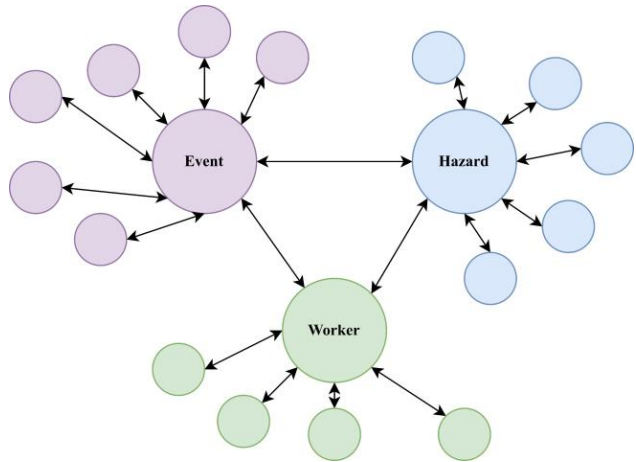

Fig. 9. Schematic diagram of a risk knowledge graph.

## 3 Application

### 3.1 CSTR description

This study selects the process system of CSTR as the basic simulation object, and the corresponding assumptions directly adopt the classic model predictive control (MPC) case demonstrated by MATLAB (Bemporad et al., 2021). The simulation to obtain process data is conducted in the software MATLAB & Simulink 2022a, and the variables are shown in Table 2.

Table 2: CSTR variable and parameter.

| Variable | Description | Initial | Equilibrium | Range |
|---|---|---|---|---|
| $u_1$ | $C_{Af}$, the concentration of the reagent in the inlet feed stream, measured in $kmol/m^3$ | 10 | 10 | - |
| $u_2$ | $T_f$, the temperature of the inlet feed stream, measured in K | 300 | 300 | $N(300,1)$ |
| $u_3$ | $T_c$, the temperature of the jacket coolant, measured in K | 292 | 299 | [276, 322] |
| $x_1$ | $C_A$, the concentration of the reagent in the reactor, measured in $kmol/m^3$ | 8.5 | 2 | [0, 10] |
| $x_2$ | $T$, the temperature in the reactor, measured in K | 311 | 373 | [310, 390] |

## 3.2 Step 1: Proactive safety actions to generate safety data

### 3.2.1 Numeric data

This study simulated the reaction in CSTR within 0-1000s, built a CSTR system in MATLAB/Simulink, and tuned the MPC to make it work normally. At 200s, the temperature sensor of the inlet feed malfunctioned, indicating that the feed temperature was too low, therefore, the upstream heater was automatically started, causing the feed temperature to be higher than the setpoint value. The simulation is by introducing a slope signal with a slope of 0.02 to represent this fault (Fig. 10).

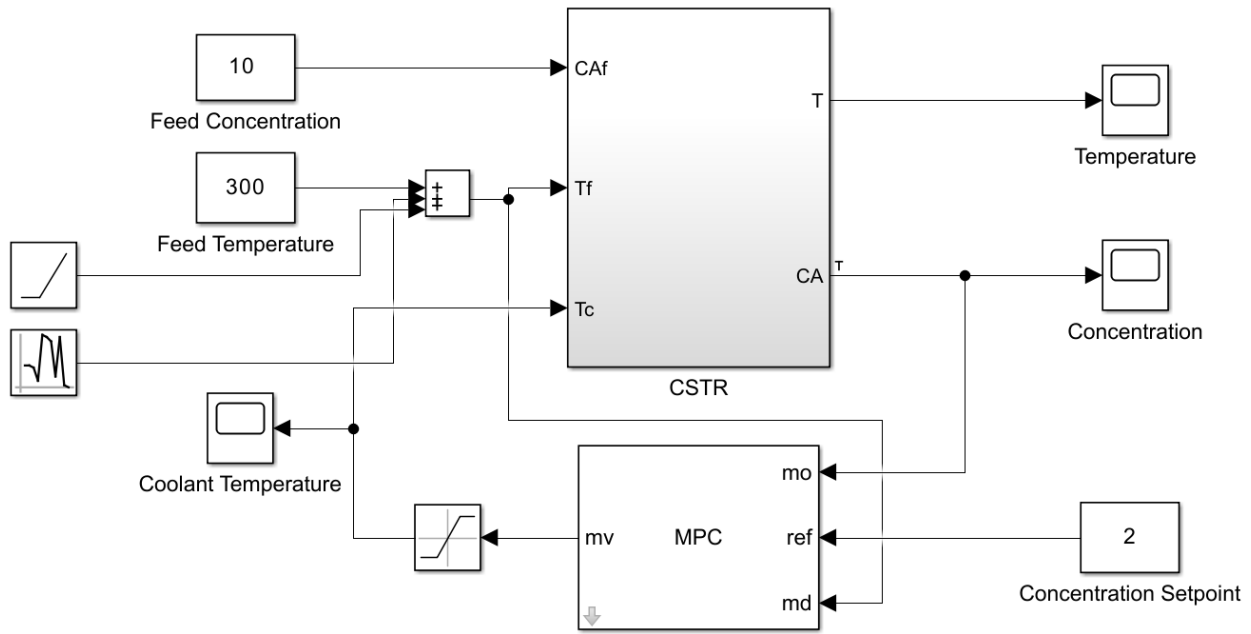

Fig. 10. Simulink Model of the CSTR with the fault.

Moreover, the operator received the alarm of the tank temperature at 550s, which was higher than the setpoint. After check, the operator also noticed that the tank concentration was lower than its setpoint, and the coolant temperature had been automatically adjusted to the minimum by the MPC. The operator's first reaction was to lower the tank temperature as soon as possible to ensure that the tank concentration returned to normal, therefore, the operator opened the cooling water inlet valve to increase the flow. In 550s-1000s, the operator's emergency response was ineffective. By simulating the above process, numerical data of three variables can be obtained (Table 3): the coolant temperature $T_c$, the tank temperature $T$, and the tank concentration $C_A$.

Table 3: Sample of collected numerical data.

| Time (s) | Coolant temperature | Tank temperature | Tank concentration |
|---|---|---|---|
| 1 | 300.36 | 373.15 | 1.9981 |
| 2 | 298.98 | 373.16 | 2.0026 |
| 3 | 299.57 | 373.05 | 1.9988 |
| 4 | 295.47 | 373.23 | 1.9985 |
| 5 | 302.74 | 373.05 | 1.9997 |

### 3.2.2 Textual data

Due to the study's reliance on simulating a system rather than employing a real setup, it is achievable to simulate the process data above, while it is challenging to propose the textual data with real operations. Therefore, in this case study, the authors present three ways to collect the textual data. First, the above simulation process has records from MPC and the operator has an operation log. Thus, an event log could be generated during the operation (Table 4).

Table 4: Example of operation event log

| Time (s) | Event Description |
|---|---|
| 200 | MPC record: Feed temperature increased. |
| 400 | MPC record: Reactor temperature maintained within normal range despite fault. |
| 550 | Alarm for tank temperature exceeding setpoint. |
|  | MPC record: coolant temperature automatically adjusted to minimum by MPC. |
|  | Operator opened cooling water inlet valve to increase flow for lowering tank temperature. |

Second, the daily safety inspection record is another source of textual data (Table 5).

Table 5: Example of daily safety inspection.

| Check item | Check result |
|---|---|
| 1. Equipment Inspection | All equipment visually inspected and found in good condition. |
| - Is the reactor vessel inspected for any damage or leaks? | No visible damage or leaks observed. |
| - Are the stirrer and its components inspected for proper functioning? | Stirrer motor and blades checked; no issues found. |
| - Are pumps checked for leaks and proper operation? | Pumps inspected; no leaks detected. |
| - Are valves inspected for leaks and proper operation? | Valve functionality verified; no leaks found. |
| - Are all instruments calibrated and functioning correctly? | Instruments calibrated; all functioning properly. |
| 2. Emergency Shutdown Procedures | Emergency shutdown procedures established and trained. |

Third, the structural hazard identification presents as textual data, for example, the HAZOP analysis sheet. This study needs to establish a risk knowledge graph, therefore, in this step, the author collected adequate information. This study searched the HAZOP analysis of CSTR on the public Internet and electronic papers/books from the libraries of the University of Alberta. The reason of choosing HAZOP as an example is that it exhibits structural knowledge by tables and the techniques has been further developed (Single et al., 2019), while the utilization in practice of HAZOP sheet is still unsatisfactory. As a result, 47 documents of HAZOP sheet that are highly correlated with the circulated topic and presented in good forms are collected (Table 6).

Table 6: Example of HAZOP sheet.

| Deviation | Causes | Consequences | Safeguards | Recommendations |
|---|---|---|---|---|
| No | 1. Pump failure<br>2. Manual valve closed<br>3. Closed block valve anywhere in water piping | 1. Potential pipe failure<br>2. Potential slip hazard | 1. Valves are visible from the control room<br>2. Standard Operating Procedure<br>3. Flowing liquid from tanker makes an audible sound | 1. Consider modifying the Standard Operating Procedure to keep the containment isolation valve closed<br>2. Install flow meters |
| Low | 1. Line Blockage<br>2. Failure of a flexible coupling at one of the tank connections | Prolonged treatment process | 1. Clear blockage<br>2. Increase pump pressure | 1. Install flow meters<br>2. Consider relocating tank isolation valves from the piping to directly on the tank, upstream of the flexible coupling |

### 3.3 Step 2: Data-driven approaches to mine safety data

#### 3.3.1 Data mining

The authors conducted exploratory data analysis of the collected numerical data. Since the trend and standard deviation clearly indicate the abnormality as shown in Fig. 11, the authors stopped the data mining process to use more advanced machine learning techniques. The S-chart, which is also one fundamental technique in data-driven fault diagnosis methods was applied. The charts indicate the fault may start in the 200s, and the abnormal situation started in the 550s.

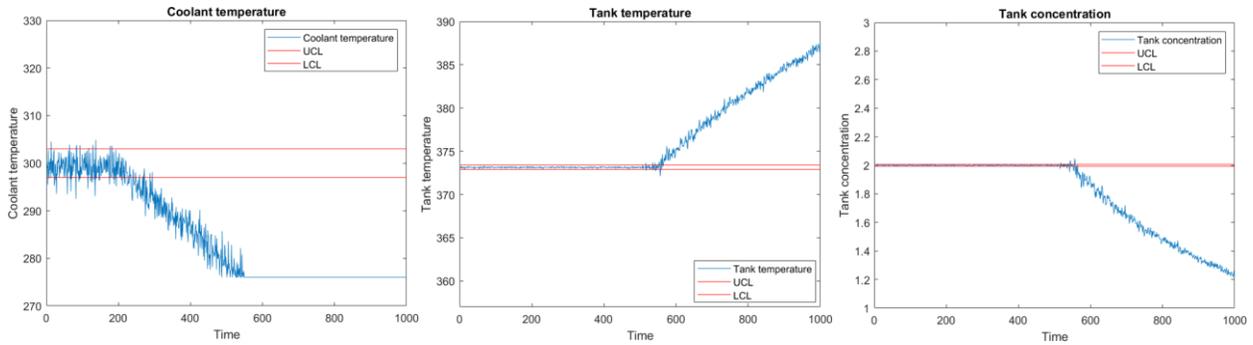

Fig. 11. Numerical data obtained by simulation.

### 3.3.2 Text mining

This study used hybrid techniques of text mining to find all tasks and events. Specifically, the authors wrote programs in Python Jupyter Notebook to analyze the collected documents, and the major steps were

i. Used pdfminer.six to extract text;
ii. Used spaCy for named entity recognition;
iii. Defined the event entity pattern as "who, when, where, what, why, how" to extract the event and its description; similarly, defined hazard pattern as "where, what";
iv. Manually corrected the acquired events.

## 3.4 Step 3: Knowledge-driven approaches to depict risk graph

### 3.4.1 Establish a risk knowledge graph

From the text mining, the collected documents imported 176 nodes (events, hazards, worker roles), and 260 "head-tail-relation" are constructed. With the hybrid approach suggested above, the authors step by step establish a risk knowledge graph for the CSTR (Fig. 12).

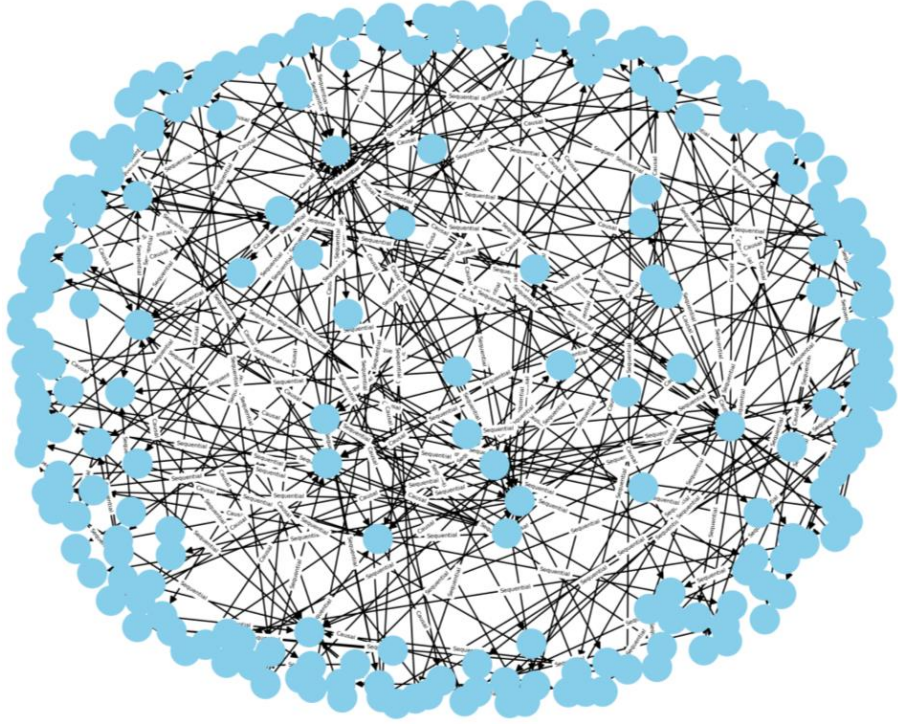

Fig. 12. The constructed risk knowledge graph for CSTR.

### 3.4.2  Examine the fault and action in the risk knowledge graph

For the simulation scenario, in 200-550s, the operator could utilize the constructed risk knowledge graph to address the situation. An automatic query based on keywords is constructed here. When the event with keywords "tank temperature" and "high" appear, the sub-knowledge graph shown in Fig. 13 is automatically extracted.

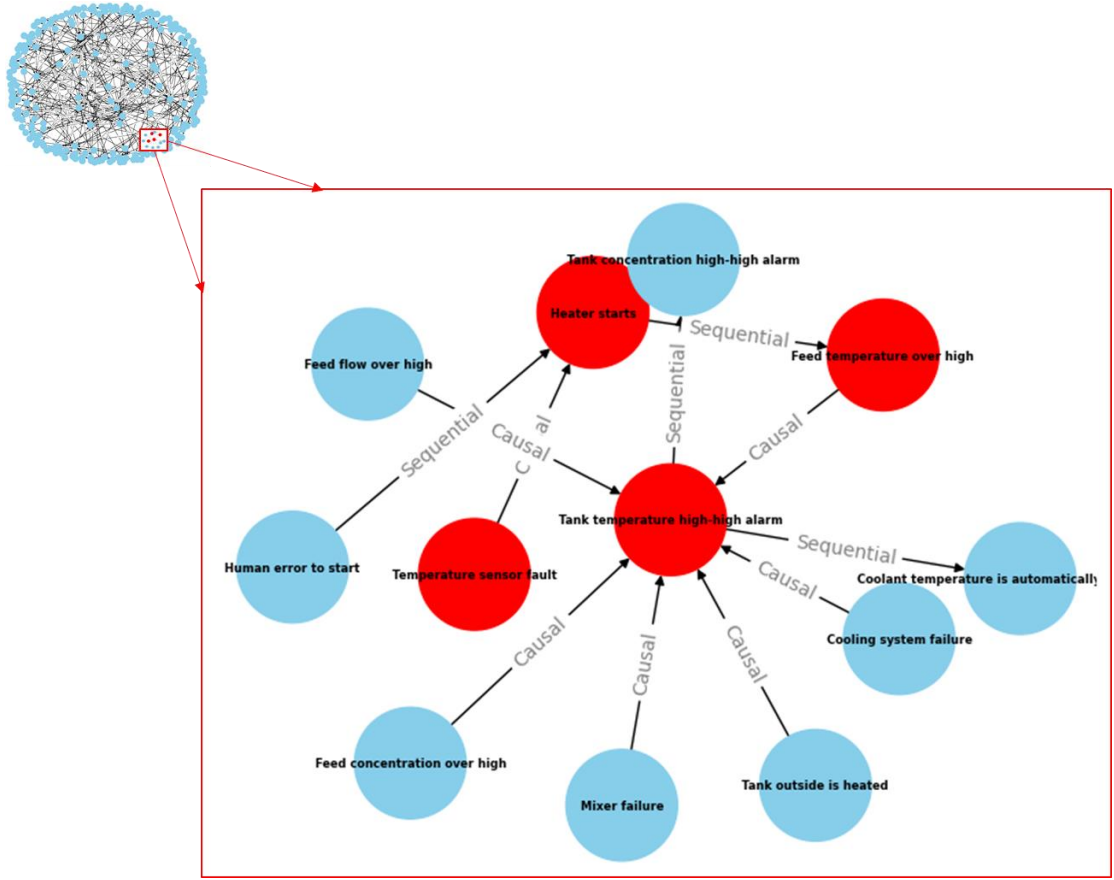

Fig. 13. Sub-knowledge graph for the simulated scenario.

Therefore, a coherent sequence of events emerges: the malfunction of the temperature sensor prompted the activation of the upstream heater, leading to an elevated feed temperature. Consequently, this elevated feed temperature resulted in an increase in the tank temperature. Thus, this interpretation helps the operator avoid opening the coolant valve by mistake and instead turn off the heater to solve the emergency.

## 4 Discussion

For the concept of proactive safety, scholars may have various understandings. Thus, this study avoids elaborating on the concept description, but shows readers its mechanism through modeling and step-by-step case analysis. Generally, this methodology and case study describes an application scenario for knowledge graphs to readers. It is the role of using a large amount of accumulated safety knowledge through a virtual AI agent to work in parallel with the operator to provide knowledge consultation and support (Fig. 14).

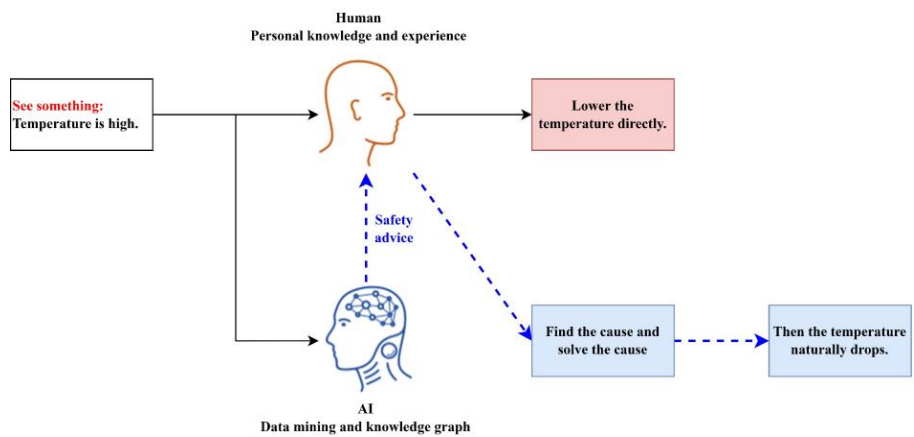

Fig. 14. A parallel decision-making auxiliary mechanism.

For the demonstration, the proposed methodology can build a series of automated processes and application software, which requires huge follow-up works and commercial applications. In this study, the authors strive to demonstrate the implementation process through MATLAB/Simulink and Python/Jupyter Notebook, and CSTR is also the most basic chemical system. This may lead to gaps with practice and practical applications, but this does not affect the effectiveness of interpretation for the proactive safety model.

Looking to the future, operators can not only interact through the knowledge graph to obtain solutions to problems, understand the causes and consequences of incidents, but also obtain knowledge-based safety checklists and take more effective preventive measures. The current rapid development of large language models may be applied to chemical engineering processes in the future. This will be a new risk and challenge, and the knowledge graph it relies on will also have richer application scenarios.

There is still potential for enhancement in this study. Firstly, further validation and refinement of the model are needed across a range of industrial domains to assess its generalizability and effectiveness in diverse safety contexts. Additionally, ongoing advancements in AI and data analytics technologies offer opportunities to enhance the capabilities of the proactive safety model. Furthermore, the scalability and interoperability of the model should be addressed to facilitate its adoption in large-scale industrial operations and heterogeneous data environments. This may involve the development of standardized data formats and interoperability protocols to enable seamless integration with existing safety management systems. Overall, continued research efforts are essential to further develop and refine the proactive safety model, ultimately contributing to the advancement of safety management practices and the prevention of workplace accidents and incidents.

## 5  Conclusions

This study proposed the model of proactive safety, integrating both data-driven and knowledge-driven approaches, leveraging advanced techniques such as data mining, natural language processing, and knowledge graph representation. This methodology represents a significant advancement in safety management practices, as it enables the transformation of raw safety data into actionable knowledge for proactive risk identification and mitigation. The application of the proactive safety model to a continuous stirred tank reactor (CSTR) scenario provides valuable insights into its practical utility and effectiveness in real-world safety contexts. The case study demonstrates the model's ability to identify and analyze safety issues proactively using both numerical and textual data sources.

Nevertheless, the proposed hybrid proactive safety model represents a prototype solution aimed at addressing longstanding challenges in safety management. It is important to acknowledge that the scenarios and data presented in this study are relatively simplified, and the methodologies employed may idealize certain aspects of real-world applications. However, despite these limitations, the model serves as a promising framework for transforming scientific research into practical applications.

## ACKNOWLEDGMENTS

The author gratefully acknowledges the financial support provided by the Natural Sciences and Engineering Research Council of Canada (NSERC) of the alliance grant- Federated Platform for Construction Simulation (572086-22).